\documentstyle[prd,aps,preprint,tighten,epsfig]{revtex}

\begin{document}

\draft

\title{One-loop renormalization group equations of the neutrino mass matrix in the triplet seesaw model}

\author{{\bf Wei Chao} \thanks{E-mail: chaowei@mail.ihep.ac.cn},
~ {\bf He Zhang} \thanks{E-mail: zhanghe@mail.ihep.ac.cn}}

\address{
CCAST (World Laboratory), P.O. Box 8730, Beijing 100080,
China \\
and
Institute of High Energy Physics, Chinese Academy of Sciences, \\
P.O. Box 918, Beijing 100049, China}

\maketitle

\begin{abstract}
Within the framework of the standard model plus one heavy Higgs
triplet, we derive a full set of one-loop renormalization group
equations of the neutrino mass matrix and Higgs couplings in both
full and effective theories. The explicit RGEs of neutrino masses,
flavor mixing angles and CP-violating phases are also obtained,
and their non-trivial running behaviors around the Higgs triplet
mass threshold are numerically illustrated.
\end{abstract}

\newpage

\section{Introduction}

To understand the origin of fermion masses and flavor mixing is
crucial and essential in modern particle physics. Current neutrino
experiments have provided very convincing evidence that neutrinos
are massive and lepton flavors are
mixed\cite{exp1,exp2,exp3,exp4,exp5}. However, it remains unclear
why the neutrino masses are quite suppressed compared to the other
fermion masses. Among various possible models, the seesaw
mechanism\cite{SS} should be one of the most favorite mechanisms
to explain the unnatural smallness of the neutrino mass scale. In
the usual Type-I seesaw mechanism, heavy right-handed Majorana
neutrinos are introduced to generate the light neutrino masses.
The lepton-flavor-violating and lepton-number-violating processes,
which are forbidden in the standard model (SM), can also take
place through the exchanges of heavy right-handed neutrinos.

Besides the Type-I seesaw scenario, the triplet seesaw
mechanism\cite{triplet,first}, which extends the SM with one Higgs
triplet $\xi=(\xi^{++},\xi^{+},\xi^{0})^{T}$ with hypercharge $1$,
gives another possible solution to the tiny light neutrino masses.
Such a Higgs triplet can be introduced in many Grand Unified
Theories (GUTs) or $SU(2)^{}_{L}\times SU(2)^{}_R$ theories. Note
that, in general the triplet seesaw model is more predictive than
the usual Type-I seesaw model, since there is no unknown
right-handed Majorana neutrino mass matrix. The minimal version of
the triplet seesaw models contains only one triplet besides the SM
particles. That is very different from the Type-I seesaw model, in
which at least two heavy right-handed Majorana neutrinos are
introduced to give rise to at least two massive left-handed
neutrinos\cite{GZZ}.

Taking into account that most of the realistic triplet seesaw
models are built at some energy scales $M^{}_{\rm NP}$ much higher
than the typical electroweak scale $M^{}_Z=91.2~{\rm
GeV}$\cite{PDG}, it is very meaningful to consider the radiative
corrections to the neutrino mixing parameters. In the energy
scales below the seesaw scale, the running of the dimension-5
operator has been considered by many authors\cite{running} in the
Type-I seesaw framework, and it has been proved that there is no
remarkable corrections in the SM or minimal supersymmetric
standard model (MSSM) with small $\tan\beta$. However, in the
energy scales above the triplet seesaw scale, the renormalization
group equations (RGEs) are quite different from those in the low
energy effective theory, and the RGE analyses are still lacking in
that energy scale. Since the RGE running may bring significant
corrections to the physical parameters, it is very important for
both model building and phenomenology to study the running effects
from the GUT scale ($M^{}_{\rm GUT}=10^{16} ~ {\rm GeV}$) to the
seesaw scale.

In this letter, we derive a full set of one-loop RGEs in the
framework of the SM extended with one heavy Higgs triplet. The
$\beta$-functions of Yukawa couplings are calculated in detail.
The RGEs of the couplings between the triplet and doublet Higgs
are also obtained. Analytical and numerical analyses based on our
formulae are also given for illustration. We show that there may
be sizable radiative corrections to the mixing parameters when the
triplet Higgs is involved.

The letter is organized as follows: In section II, the basic
concepts of the triplet seesaw model is briefly discussed. In
section III, we present the $\beta$-functions of Yukawa and Higgs
couplings in the model. Section IV is dedicated to the numerical
analyses of the running effects on mixing parameters. Finally, a
summary is given in section V.
\section{the triplet seesaw model}
The full Lagrangian of the triplet seesaw model is given by
\cite{first}:
\begin{eqnarray}
{\cal L}_{\rm full}^{}={\cal L}_{\rm SM}^{}+{\cal L}_\xi^{} \ ,
\end{eqnarray}
where the first part represents the SM Lagrangian, and the second
part contains the interactions involving the Higgs triplet. The
most general form of ${\cal L}^{}_\xi$ is
\begin{eqnarray}
{\cal L}^{}_\xi & = & (D^{}_\mu \xi)^\dagger (D^{}_\mu \xi)
-M_{\xi}^2(\xi^\dagger \xi)-{1\over 4}\lambda_\xi^{}(\xi^\dagger
\xi)^2-\lambda^{}_\phi (\xi^\dagger \xi) (\phi^\dagger
\phi)-\lambda_{C}^{}(\xi^T_{}\hat{C}\xi)^\dagger_{}(\xi^T_{}\hat{C}\xi)
 \nonumber \\ &&  -\lambda_T^{}(\xi^\dagger_{}t_i^{}\xi)
(\phi^\dagger_{}{\tau_i^{}\over 2}\phi)-{1\over 2}\left[
(Y_\xi^{})_{fg}^{}\overline{ l^C_{L}}^f \varepsilon \Delta
l_{L}^{g}+\lambda_H^{} M_\xi^{} \tilde{\phi}^T \varepsilon \Delta
\tilde{\phi}+{\rm h.c.} \right ] \ ,
\end{eqnarray}
where $\phi$ is the SM Higgs doublet with
$\tilde{\phi}=i\tau^{}_2\phi^{*}$ and $\Delta$ is a $2\times 2$
representation of the Higgs triplet field\cite{lagranglian}:
\begin{eqnarray}
\Delta=\left (\matrix{\displaystyle {\xi^+/ \sqrt{2}} &
\displaystyle -\xi^{++}_{}\cr \displaystyle \xi^0_{}&
\displaystyle -{\xi^+ / \sqrt{2}}\cr}\right) \ .
\end{eqnarray}
In Eq. (2), $f$ and $g$ are generation indices, and summation over
repeated indices is implied. $t^{}_i$ are the three dimensional
representations of the Pauli matrices
\begin{eqnarray}
t^{}_1={1\over \sqrt{2}}\left( \matrix{ 0&1&0\cr
1&0&1\cr0&1&0\cr}\right) \ , \ \ \ \ \  t^{}_2={1\over
\sqrt{2}}\left( \matrix{ 0&-i&0\cr i&0&-i\cr0&i&0\cr}\right) \ , \
\ \ \ \  t^{}_3=\left( \matrix{ 1&0&0\cr 0&0&0\cr0&0&-1\cr}\right)
\ ,
\end{eqnarray}
and $\hat{C}$ is defined as\cite{C}:
\begin{eqnarray}
\hat{C}=\left( \matrix{ 0 & 0 & -1 \cr 0 & 1 & 0 \cr -1 & 0 & 0 }
\right) \ .
\end{eqnarray}
The covariant derivative $D^{}_\mu$ reads
\begin{eqnarray}
D^{}_\mu \xi\equiv\partial^{}_\mu \xi+ig^{}_{1} Y
B_\mu^{}\xi+ig^{}_2 t \cdot W_\mu^{} \xi \ .
\end{eqnarray}

As already mentioned in the first section, in order to generate
tiny light neutrino masses, the mass scale of the Higgs triplet
$M^{}_\xi$ should be much higher than the typical electroweak
scale $M_Z^{}$. In this letter, we take $M^{}_\xi = 10^{10}~ {\rm
GeV}$. In the energy scale $\mu \gg M^{}_{\xi}$, the full
Lagrangian is taken into account. In the low energy limit $\mu \ll
M^{}_{\xi}$, we should use the effective theory by integrating out
the heavy triplet field. The effective Lagrangian can be defined
by\cite{integrating} :
\begin{eqnarray}
\exp\left\{i\int d^4x{\cal L}^{}_{\rm{eff}}(x)\right\} & \equiv &
\int \prod_i^3 {\mathcal D} \xi^{}_i {\mathcal D} \xi^{\dagger}_i
~{\exp}\left\{ i\int d_{}^4x{\cal L}_{\rm full}^{}(x) \right\}
\nonumber \\
& = &\exp \left\{ i\int d_{}^4x{\cal L}^{}_{\rm SM}(x)
\right\}\int \prod_i^3 {\mathcal D} \xi^{}_i {\mathcal D}
\xi^{\dagger}_i ~\exp\left\{i\int d_{}^4x{\cal L}^{}_\xi
(x)\right\} \ ,
\end{eqnarray}
where ${\mathcal D}\xi^{}_i$ stands for functional integration
over $\xi^{}_{i}$. Starting from Eq. (7), we can calculate the
one-loop level effective Lagrangian by using the steepest-descent
method to integrate out the heavy scalar. Keeping only terms of
order ${\mathcal O}(1/M^2_{\xi})$ through the whole calculation
and neglecting all the operators with higher inverse powers of
$M^{}_\xi$, we may get the effective operators:
\begin{eqnarray}
{\cal L}_{4-\rm Higgs}&=&-{1\over 4}
\lambda_H^\dagger\lambda_H^{}(\phi^\dagger_{}\phi)^2_{} \ ;  \nonumber \\
{\cal L}_{\nu-\rm mass}^{}&=&-{1\over
4}{\lambda_H^{}(Y_\xi^{})^{}_{fg}\over
M_\xi^{}}(\overline{l^C_L}^f \varepsilon\phi )(l^{g}_{L}
\varepsilon\phi )+{\rm h.c.} \ ; \nonumber \\
{\cal L}_{\rm 4-Fermi}&=&-{1\over
8}{(Y_\xi^\dagger)_{mn}(Y_\xi)_{fg}\over
M_\xi^2}\left[(\overline{l_L^m}\cdot
l_L^g)(\overline{l_L^C}^f\cdot{l_L^C}^n)+({\overline{l_L^C}}^f\cdot\overline
{l_L^m}^T)({l_L^g}^T\cdot{l_L^C}^n)\right ] \ ,
\end{eqnarray}
where $m,n,f$ and $g$ run over $1,2,3$. ${\cal L}^{}_{\rm Higgs}$
has the same form as the Higgs self-coupling in the SM. The four
fermion interaction ${\cal L}^{}_{\rm 4-Fermi}$ may contribute to
the lepton flavor violating processes. However, such processes
should be strongly suppressed due to the heavy mass of the Higgs
triplet. Thus we will neglect this four fermion coupling in the
following calculations. ${\cal L}^{}_{\nu-\rm mass}$ is in
proportion to the Majorana neutrino mass matrix. In analogy to the
Type-I seesaw model, we can also define the effective dimension-5
operator
\begin{eqnarray}
{\cal L}^{}_{\nu-\rm mass} = -{1\over4}
\kappa^{}_{fg}(\overline{l^C_L}^f \varepsilon\phi )(l^{g}_{L}
\varepsilon\phi )+{\rm h.c.}\ ,
\end{eqnarray}
where $\kappa=\lambda^{}_H Y^{}_{\xi}/M^{}_\xi$. After spontaneous
symmetry breaking, neutrinos acquire masses and the neutrino mass
matrix is given by $M^{}_\nu=Y^{}_{\xi}\left< \Delta ^0\right>$
with $\left< \Delta^0 \right>=\lambda^{}_H v^2/M^{}_\xi$. Here
$v\simeq 174 ~ {\rm GeV}$ denotes the Higgs vacuum expectation
value. Since $M^{}_\xi \gg v$, the mass scale of neutrinos is then
suppressed, and this is the so-called triplet seesaw mechanism.

\section{calculations of the Beta-functions}

In our calculations, we always use the dimensional regularization.
For the one-loop wavefunction renormalization constants $Z$ above
the seesaw scale, we find that
\begin{eqnarray}
\delta Z^{}_{l^{}_L}&=&-{1\over 16\pi^2}\left({3 \over 2}
Y^\dagger_\xi Y^{}_\xi+Y_e^\dagger Y_e^{} + {1 \over 2}
g_1^2\xi_B^{} + {3 \over 2} g_2^2\xi^{}_W\right) {1\over
\varepsilon} \ , \nonumber
\\ \delta Z^{}_\phi & = & -{1\over 16\pi^2} \left
[2{\rm{Tr}}\left(Y_e^\dagger Y_e^{}+3 Y_d^\dagger Y_d^{}+3
Y_u^\dagger Y_u^{}\right )+ {1 \over 2}g_1^2(\xi_B^{}-3)+{3 \over
2}g_2^2(\xi_W^{}-3)\right]{1\over \varepsilon} \ ,  \nonumber  \\
\delta Z_\xi^{}&=&-{1\over 16\pi^2}\left [{\rm {Tr}}(Y_\xi^\dagger
Y_\xi^{})+2g_1^2(\xi_B^{}-3)+4g_2^2(\xi_W^{}-3)
\right]{1\over\varepsilon} \ .
\end{eqnarray}
For the vertex renormalization constants and the Higgs masses, we
obtain
\begin{eqnarray}
\delta Z_{Y_{\xi}^{}}&=&{1\over
16\pi^2}\left[{1\over2}g_1^2(3-3\xi_B^{})+{1\over2}g_2^2(3-7\xi_W^{})
\right]{1\over \varepsilon} \nonumber \ ,  \\
\delta\lambda_H^{}&=&-{1\over16\pi^2}\left [{3\over2}g_1^2\xi_B^{}
+{7\over2}g_2^2\xi_W^{}\right]{1\over \varepsilon} \ , \nonumber \\
\delta m^2_\phi &=&{1\over16\pi^2}\left[3 \lambda m^2_\phi +
6\lambda_\phi^{}m_\xi^2+3\lambda^{\dagger}_H\lambda^{}_H m^2_\xi -
{1 \over 2}(g_1^2\xi_B^{}+ 3 g_2^2\xi_W^{})m^2_\phi \right]{1\over
\varepsilon}
\nonumber \ , \\
\delta m_\xi^2&=&{1\over16\pi^2}\left[ 4\lambda_\xi^{} m_\xi^2+ 2
\lambda^{}_C m^2_\xi+
4\lambda_\phi^{}m_\phi^2+\lambda^{\dagger}_H\lambda^{}_H
m^2_\xi-2(g_1^2\xi_B^{}+2g_2^2\xi_W^{})m_\xi^2 \right]{1\over
\varepsilon}  \ .
\end{eqnarray}
The counterterms of the Higgs couplings $\lambda^{}_\xi$,
$\lambda^{}_C$, $\lambda^{}_\phi$ and $\lambda^{}_T$ have also
been calculated,
\begin{eqnarray}
\delta\lambda_\xi^{}&=&{1\over
16\pi^2}\left\{6\lambda_\xi^2+2\lambda_T^2+8\lambda_\phi^2+4\lambda_{C}^2+4\lambda_\xi^{}\lambda_{C}^{}
-4g_1^2\lambda_\xi^{}-8g_2^2\lambda_\xi^{}+24g_1^4+72g_2^4 \right. \nonumber \\
&&\left .+48g_1^2g_2^2+{\rm Tr}[(Y_\xi^{\dagger}
Y_\xi^{})^2]\right\}{1\over \varepsilon} \nonumber  \ , \\
\delta\lambda_{C}^{}&=&{1\over16\pi^2}\left\{3\lambda_{C}^2
+6\lambda_\xi^{}\lambda_{C}^{}-2\lambda_T^2
-4g_1^2\lambda_{C}^{}-8 g_2^2 \lambda_{C}^{}-48g_1^2g_2^2 -2{\rm
Tr}[(Y_\xi^{\dagger}Y_\xi^{})^2]\right\}{1\over \varepsilon} \nonumber  \ ,\\
\delta{\lambda_\phi^{}}&=&{1\over 16\pi^2}\left(4\lambda_\phi^2
+2\lambda_T^2+4\lambda_\xi^{}\lambda_\phi^{}+2\lambda_{C}^{}
\lambda_\phi^{}+3\lambda_\phi^{}\lambda-{5\over
2}g_1^2\lambda_\phi^{}-{11\over2}g_2^2\lambda_\phi^{}+3g_1^4+9g_2^4\right){1\over
\varepsilon} \nonumber \ , \\
\delta{\lambda_T^{}}&=&{1\over16\pi^2}\left(8\lambda_\phi^{}\lambda_T^{}+
\lambda_\xi^{}\lambda_T^{}-2\lambda_{C}^{}\lambda_T^{}+\lambda\lambda_T^{}
-{5\over2}g_1^2\lambda_T^{}-{11\over2}g_2^2\lambda_T^{}+12g_1^2g_2^2\right)
{1\over \varepsilon} \ .
\end{eqnarray}

By using the counterterms calculated above and the technique
described in \cite{running}, we obtain the $\beta$-functions
($\beta^{}_X=\mu{{\rm d}\over {\rm d} \mu} X$) of Yukawa couplings
and $\lambda_H^{}$:
\begin{eqnarray}
16\pi^2_{}\beta_{Y_\xi}^{}&=&Y_\xi^{}\left[{3\over
4}(Y_\xi^\dagger Y_\xi^{})^{}_{}+{1\over 2}(Y_e^\dagger
Y_e^{})\right]+\left[{3\over 4}(Y_\xi^\dagger
Y_\xi^{})^T_{}+{1\over 2}(Y_e^\dagger
Y_e^{})^T_{}\right]Y_\xi^{} \nonumber \\
\nonumber &&+ {{1\over 2}}\left[{ \rm Tr}\left(Y_\xi^\dagger
Y_\xi^{}\right) -\left(3g_1^2+9g_2^2\right) \right] Y_\xi^{} \ ,
\nonumber \\
16\pi^2\beta_{Y_e^{}}&=&Y_e^{}\left[{3\over 4}Y_\xi^\dagger
Y_\xi^{}+{3\over 2}Y_e^\dagger Y_e^{}+{\rm Tr}(Y_e^\dagger
Y_e^{}+3Y_u^\dagger Y_u^{}+3Y_d^\dagger Y_d^{})-{15\over 4}g_
1^2-{9\over 4}g_2^2\right] \nonumber,\\
16\pi^2_{}\beta_{\lambda_H}&=&\lambda_H^{}{ \rm Tr}\left( {1\over
2} Y_\xi^\dagger Y_\xi+2 Y_e^\dagger Y_e+6 Y_u^\dagger Y_u+6
Y_d^\dagger Y_d\right)+{1\over 2}\lambda_H^{}(-9 g_1^2-21 g_2^2) \
,
\end{eqnarray}
and the anomalous dimensions ($\gamma^{}_m=-{1\over m }\mu{{\rm d}
m \over {\rm d} \mu} $) of the Higgs masses:
\begin{eqnarray}
16 \pi^2 \gamma_{m^{}_\phi} & = &- \left[ {3 \over 2} \lambda +
{\rm Tr} \left ( Y^{\dagger}_e Y^{}_e + 3 Y^{\dagger}_u Y^{}_u + 3
Y^{\dagger}_d Y^{}_d \right ) \right ] +{3\over 4}g_1^2+{9\over
4}g_2^2- 3 ( \lambda^{}_\phi + {1\over 2}\lambda^{\dagger}_H
\lambda^{}_H )
{m^2_\xi\over m^2_\phi} \ , \nonumber \\
16 \pi^2 \gamma_{m^{}_\xi} & = &- \left[ 2\lambda_\xi^{}+
\lambda^{}_C +{1\over 2}\lambda_H^\dagger\lambda_H^{}+ {1\over
2}{\rm{Tr}}\left(Y_\xi^\dagger
Y_\xi^{}\right)-3g_1^2-6g_2^2\right]- 2 \lambda_\phi^{} {m_\phi^2
\over m^{2}_\xi} \ .
\end{eqnarray}
It should be noticed that in the triplet seesaw model the mass of
the Higgs doublet suffers from the so-called $ hierarchy$
$problem$, which can be prevented in some other supersymmetric
models. We also calculate the $\beta$-functions of Higgs
couplings:
\begin{eqnarray}
16\pi^2\beta_{\lambda}&=&6\lambda^2_{}+12\lambda_\phi^2+2\lambda_T^2
-3\lambda\left(g_1^2+3g_2^2\right)+3g_2^4 + {3\over
2}\left(g_1^2+g_2^2\right)^2_{}\ \nonumber \\&&+4\lambda{\rm
Tr}\left(Y_e^\dagger
Y_e^{}+3Y_u^\dagger Y_u^{}+3Y_d^\dagger Y_d^{}\right) \nonumber \\
&&-8{\rm Tr}\left[(Y_e^\dagger Y_e^{})^2+3(Y_u^\dagger Y_u^{})^2
+3(Y_d^\dagger Y_d^{})^2\right]\nonumber \\
16\pi^2\beta_{\lambda_\xi}&=&2\lambda_T^2+8\lambda_\phi^2+4\lambda_\xi^{}
\lambda_{C}^{}+4\lambda_{C}^2+6\lambda_\xi^2
-12g_1^2\lambda_\xi^{}-24g_2^2\lambda_\xi^{}+24g_1^4 \nonumber \\
&&+72g_2^4+48g_1^2g_2^2+{\rm 4Tr}\left[(Y_\xi^\dagger Y_\xi^{}
)^2\right]+2\lambda_\xi^{}{\rm Tr}\left(Y_\xi^\dagger Y_\xi^{}\right) \nonumber\\
16\pi^2\beta_{\lambda_C}&=&3\lambda_{C}^2+6\lambda_\xi^{}\lambda_{C}^{}-
2\lambda_T^2-12g_1^2\lambda_{C}^{}-24 g_2^2
\lambda_{C}^{}-48g_1^2g_2^2 \nonumber\\ &&+2\lambda_C^{}{\rm
Tr}\left(Y_\xi^\dagger Y_\xi\right)-2{\rm Tr}\left[(
Y_\xi^\dagger Y_\xi^{})^2\right] \nonumber \\
16\pi^2\beta_{\lambda_\phi}&=&4\lambda_\phi^2+2\lambda_T^2
+4\lambda_\xi^{}\lambda_\phi^{}+2\lambda_{C}^{}\lambda_\phi^{}+3\lambda\lambda_\phi^{}
-{15\over2}g_1^2\lambda_\phi^{}-{33\over2}g_2^2\lambda_\phi^{}+9g_2^4\nonumber\\&&+3g_1^4
+2\lambda_\phi^{}{\rm Tr}\left({1\over 2}Y_\xi^\dagger
Y_\xi^{}+Y_e^\dagger Y_e^{}+3Y_d^\dagger Y_d^{}+3Y_u^\dagger
Y_u^{}\right) \nonumber
\\16\pi^2\beta_{\lambda_T}&=&8\lambda_\phi\lambda_T^{}+
\lambda_\xi^{}\lambda_T^{}-2\lambda_{C}^{}\lambda_T^{}+\lambda\lambda_T^{}-{15\over
2}g_1^2\lambda_T^{}-{33\over2}g_2^2\lambda_T^{}
+12g_1^2g_2^2\nonumber\\&&-2\lambda_T^{}{\rm Tr}\left({1\over
2}Y_\xi^\dagger Y_\xi^{}+Y_e^\dagger Y_e^{}+3Y_d^\dagger
Y_d^{}+3Y_u^\dagger Y_u^{}\right) \ .
\end{eqnarray}
The RGEs for the gauge couplings are changed in this model, and we
list the results below:
\begin{eqnarray}
16\pi^2 \beta_{g_1}^{} & = & {47\over 6} g_1^3 \ , \nonumber \\
16\pi^2 \beta_{g_2}^{} & = & -{5\over 2} g_2^3 \ , \nonumber \\
16\pi^2 \beta_{g_3}^{} & = & -{7} g_3^3 \ .
\end{eqnarray}
Since the Higgs triplet field does not couple with quarks, the
RGEs of Yukawa couplings for quarks are the same as those in the
SM, and the corresponding results can be found in the
literature\cite{RGE}.

By calculating the relevant one-loop diagrams, we obtain the
$\beta$-function of the effective operator which describes the
neutrino masses and mixing at the energy scales below the seesaw
scale:
\begin{eqnarray}
16\pi^2\beta_{\kappa}&=&-{3\over 2}\kappa(Y_e^\dagger
Y_e^{})^T_{}-{3\over 2}(Y_e^\dagger
Y_e^{})\kappa+(\lambda+\lambda_H^\dagger\lambda_H^{})\kappa-3g_2^2\kappa
\nonumber \\
&&+2{\rm Tr}\left(Y_e^\dagger Y_e^{}+3Y_u^\dagger
Y_u^{}+3Y_d^\dagger Y_d^{}\right)\kappa \ .
\end{eqnarray}
It should be mentioned that Eq. (17) has the same form as that in
the Type-I seesaw model, only up to a replacement
$\lambda\rightarrow \lambda +\lambda^{\dagger}_H\lambda^{}_H$.


\section{Applications}

To see the running behaviors of neutrino mixing parameters in the
triplet seesaw model, we carry out some numerical analyses by
using the $\beta$-functions derived above. The lepton flavor
mixing matrix, which comes from the mismatch between the
diagonalizations of the neutrino mass matrix and the charged
lepton mass matrix, is given by $V=V^{\dagger}_e V^{}_\xi$, where
\begin{eqnarray}
V^{\dagger}_e \cdot (Y^{\dagger}_e Y^{}_e)\cdot V^{}_e & = &
{\rm Diag}\left( y^{2}_e , ~ y^2_\mu , ~ y^2_\tau\right) \ , \nonumber \\
V^{T}_\xi \cdot Y^{}_\xi \cdot V^{}_\xi & = & {\rm Diag} \left(
y^{}_1 , ~ y^{}_2, ~ y^{}_3\right) \ ,
\end{eqnarray}
with ($y^{}_e$, $y^{}_\mu$, $y^{}_\tau$) and ($y^{}_1$, $y^{}_2$,
$y^{}_3$) being the eigenvalues of $Y^{}_e$ and $Y^{}_\xi$
respectively. In this letter, we adopt the following
parameterization\cite{para}:
\begin{equation}
V = \left( \matrix{ c^{}_{12}c^{}_{13} & s^{}_{12}c^{}_{13} &
s^{}_{13} \cr -c^{}_{12}s^{}_{23}s^{}_{13} - s^{}_{12}c^{}_{23}
e^{-i\delta} & -s^{}_{12}s^{}_{23}s^{}_{13} + c^{}_{12}c^{}_{23}
e^{-i\delta} & s^{}_{23}c^{}_{13} \cr -c^{}_{12}c^{}_{23}s^{}_{13}
+ s^{}_{12}s^{}_{23} e^{-i\delta} & -s^{}_{12}c^{}_{23}s^{}_{13} -
c^{}_{12}s^{}_{23} e^{-i\delta} & c^{}_{23}c^{}_{13} } \right)
\left ( \matrix{e^{i\rho } & 0 & 0 \cr 0 & e^{i\sigma} & 0 \cr 0 &
0 & 1 \cr} \right ) \; ,
\end{equation}
where $c^{}_{ij} \equiv \cos\theta_{ij}$ and $s^{}_{ij} \equiv
\sin\theta_{ij}$ (for $ij=12,23$ and $13$).

For the typical choice $\left< \Delta^0 \right> \sim {\mathcal
O}(0.1~ \rm eV)$, one can estimate that $Y^{}_\xi \sim {\mathcal
O}(1)$, which means that the relation $Y^{}_{\xi}\gg Y^{}_e$
holds, and the tiny Yukawa coupling $Y^{}_{e}$ in the RGEs of
$Y^{}_\xi$ and $Y^{}_e$ can be safely neglected in the
hierarchical mass spectrum case
\footnote{When the neutrino mass spectrum is nearly degenerate
$m^{}_1\simeq m^{}_2 \simeq m^{}_3$, such an approximation may not
be reasonable. However, in our numerical calculations, we use the
exact RGEs and do not make any approximations.}.
Then we get the approximate equations of Yukawa couplings:
\begin{eqnarray}
\mu\frac{{\rm d} Y^{}_{\xi}}{{\rm d}\mu} & \simeq &
\frac{1}{16\pi^2} Y^{}_{\xi}\left( {3 \over 2} Y^{\dagger}_{\xi}
Y^{}_{\xi} + \alpha^{}_{\xi} \right) \ , \nonumber \\
\mu\frac{{\rm d} Y^{}_{e}}{{\rm d}\mu} & \simeq &
\frac{1}{16\pi^2}Y^{}_e \left( {3\over 4} Y^{\dagger}_{\xi}
Y^{}_{\xi} + \alpha^{}_{e} \right) \ ,
\end{eqnarray}
where
\begin{eqnarray}
\alpha^{}_\xi&=&  {{1\over 2}}\left[{ \rm Tr}\left(Y_\xi^\dagger
Y_\xi^{}\right) -\left(3g_1^2+9g_2^2\right) \right] \ ,
\nonumber \\
\alpha^{}_e &=&{\rm Tr} \left (Y_e^\dagger Y_e^{}+3Y_u^\dagger
Y_u^{}+3Y_d^\dagger Y_d^{} \right )-{15\over 4}g_ 1^2-{9\over
4}g_2^2 \ .
\end{eqnarray}
Using the results above and taking into account the fact that
$m^{}_\tau
> m^{}_\mu \gg m^{}_e$, we neglect the tiny terms in proportion to
$y^{2}_{e}$ and arrive at the approximate analytical results of
three mixing angles:
\begin{eqnarray}
\mu\frac{{\rm d} \theta^{}_{12}}{{\rm d}\mu} & \simeq &-
\frac{1}{16\pi^2}\cdot {3\over 4} s^{}_{12}c^{}_{12}\Delta^{}_{21}
 \ ,  \\
\mu\frac{{\rm d} \theta^{}_{23}}{{\rm d}\mu} & \simeq &
\frac{1}{16\pi^2}\cdot \frac{3}{4}  \left \{ \left [2 \cos\delta
s^{}_{12} c^{}_{12} c^2_{23} s^{}_{13}- s^{}_{23}c^{}_{23} (
s^{2}_{12}-c^{2}_{12} s^{2}_{13} )  \right ] \Delta^{}_{21} -
s^{}_{23} c^{}_{23} c^2_{13} \Delta^{}_{32}
\right \} \ , \\
\mu \frac{{\rm d} \theta^{}_{13}}{{\rm d}\mu} & \simeq & -
\frac{1}{16\pi^2}\cdot{3\over 4} s^{}_{13}c^{}_{13}\left(
c^{2}_{12}\Delta^{}_{21} +\Delta^{}_{32} \right)  \ ,
\end{eqnarray}
where $\Delta^{}_{ij}=y^2_i-y^2_j$ with ($i,j=1,2,3$). Note that,
in deriving Eqs. (22)-(24), we have adopted the the
parametrization in Eq. (19). Such instructive expressions allow us
to do useful analyses of the running behaviors of mixing angles.
Due to the hierarchical charged lepton masses, there is in general
no enhanced factor compared with the Type-I seesaw
model\cite{running}. However, nontrivial running effects may also
be acquired from the sizable Yukawa coupling $Y^{}_\xi$. From Eqs.
(22)-(24), one can immediately conclude that the corrections to
$\theta^{}_{12}$ and $\theta^{}_{13}$ should be milder than that
to $\theta^{}_{23}$ since the right-hand sides of Eqs. (22) and
(24) are in proportion to either $\Delta^{}_{21}$ or
$\theta^{}_{13}$. In the limit $\theta^{}_{13}\rightarrow 0$ and
$\Delta^{}_{21}\rightarrow0$, we can see from Eq. (23) that
$\dot{\theta}^{}_{23}\propto -\Delta^{}_{32}$. Thus
$\theta^{}_{23}$ will get negative correction in the normal
hierarchy case. For illustration, we only show the evolution of
$\theta^{}_{23}$ with different $\lambda^{}_H(M^{}_\xi)$ in Fig.
1. We can see that its running is quite sensitive to
$\lambda^{}_H$ and a decrease of several degrees may be acquired
from the RGE evolution.

Considering the smallness of $\theta^{}_{13}$, the evolution of
the Dirac CP-violating phase $\delta$ is the same as those of two
Majorana phases ($\rho$, $\sigma$) at the leading order of
$s^{-1}_{13}$,
\begin{eqnarray}
\mu\frac{{\rm d} \delta}{{\rm d}\mu}  & \simeq &  \mu\frac{{\rm d}
\rho}{{\rm d}\mu} \simeq \mu\frac{{\rm d} \sigma}{{\rm d}\mu}
\simeq \frac{1}{16\pi^2}\cdot \frac{ 3y^2_e
(y^2_\tau-y^2_\mu)}{2(y^2_\tau-y^2_e)(y^2_\mu-y^2_e)}
\frac{\sin\delta s^{}_{12}c^{}_{12} s^{}_{23}
c^{}_{23}}{s^{}_{13}} \Delta^{}_{21} + {\mathcal
O}(\theta^{}_{13}) \ .
\end{eqnarray}
This is an interesting feature: once three CP-violating phases are
same at certain energy scale, they will keep this equality against
the RGE running. We can also see that the small
$\sin\theta^{}_{13}$ in the denominator of Eq. (25) dominates the
running of CP phases. That means a fixed point\cite{fix} should
exist for extremely tiny $\theta^{}_{13}$. As an example, we plot
the evolution of $\delta$ with different $\theta^{}_{13}$ in Fig.
2. Similar results can be obtained for two Majorana phases $\rho$
and $\sigma$.

By using Eq. (20), we obtain the RGEs of the eigenvalues of
$Y^{}_{\xi}$
\begin{eqnarray}
\mu\frac{{\rm d} y^{}_i } {{\rm d}\mu}  & \simeq &
\frac{1}{16\pi^2} \left ( {3\over2} y^3_i + \alpha^{}_\xi y^{}_i
\right ) \ ,
\end{eqnarray}
with $i=1,2,3$. Note that, for different signs of $\alpha^{}_\xi$,
the corrections to $y^{}_i$ may be either positive or negative.
However, in order to investigate the running of light neutrino
masses, one should consider the RGEs of $m^{}_\xi$ and
$\lambda^{}_H$ simultaneously. In Fig. 3, we present the typical
evolution of three light neutrino masses with
$\lambda^{}_H(M^{}_\xi) = 5\times 10^{-5}$. We can see that their
running effects are appreciable and should not be neglected. A
detailed numerical analysis of the triplet seesaw model is
worthwhile and the corresponding work will be elaborated
elsewhere.

\section{summary}

Working in the framework of the SM extended with one heavy Higgs
triplet, we have derived a full set of one-loop RGEs for lepton
Yukawa and Higgs couplings. Since the triplet seesaw model
involves more couplings than the usual Type-I seesaw models, the
results are also quite different. Analytical and numerical
analyses have been given based on the RGEs we obtained. We find
that nontrivial corrections to the mixing parameters can be
acquired and they should not be neglected in general. It provides
us a possible way to connect the experimental values of lepton
flavor mixing parameters with some high energy GUT theories. In
conclusion, our formulae are very important for both model
building and phenomenological analyses of the triplet seesaw
models.

\begin{acknowledgments}
The authors are indebted to Professor Zhi-zhong Xing for reading
the manuscript with great care and patience, and for his valuable
comments and numerous corrections. They are also grateful to S.
Zhou for useful discussions. This work was supported in part by
the National Nature Science Foundation of China.
\end{acknowledgments}


\newpage

\begin{figure}
\psfig{file=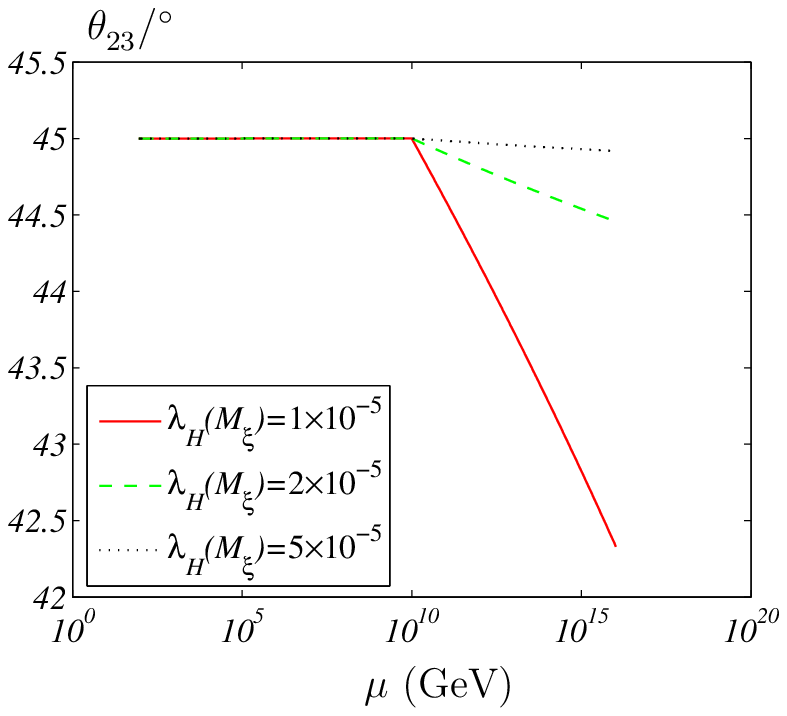, bbllx=1cm, bblly=6.0cm, bburx=11cm, bbury=16.0cm,%
width=11cm, height=11cm, angle=0, clip=0}\vspace{-2cm}\caption{The
evolution of $\theta^{}_{23}$. We take
$\delta=\rho=\sigma=90^\circ$ and $\theta^{}_{13}=0.01^\circ$ at
the scale $\mu=M^{}_{Z}$.}
\end{figure}
\begin{figure}
\vspace{-0cm}
\psfig{file=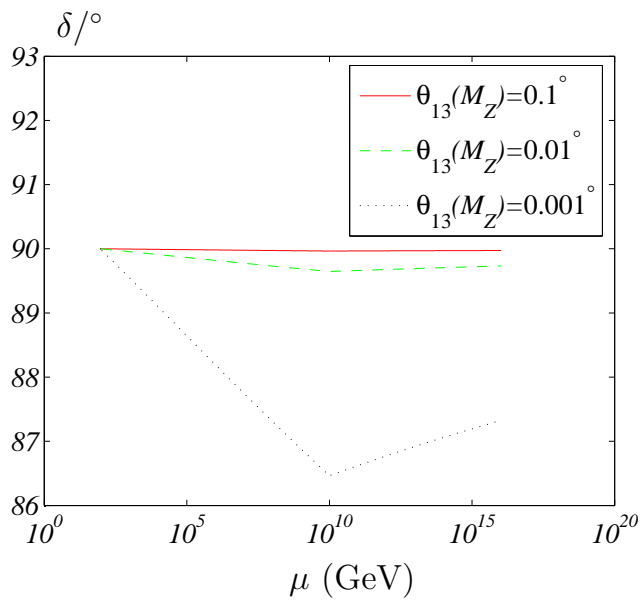, bbllx=1cm, bblly=6.0cm, bburx=11cm, bbury=16.0cm,%
width=11cm, height=11cm, angle=0,
clip=0}\vspace{-2cm}\caption{Examples of the evolution of the
Dirac CP-violating phase $\delta$. We take $\delta=90^\circ$ and
$m^{}_{1}=0.01~{\rm eV}$ at the $M^{}_{Z}$ scale. We also take
$\lambda^{}_H(M^{}_Z)=5\times10^{-5}$. Similar results can be
obtained for two Majorana phases $\rho$ and $\sigma$.}
\end{figure}

\newpage

\begin{figure}
\psfig{file=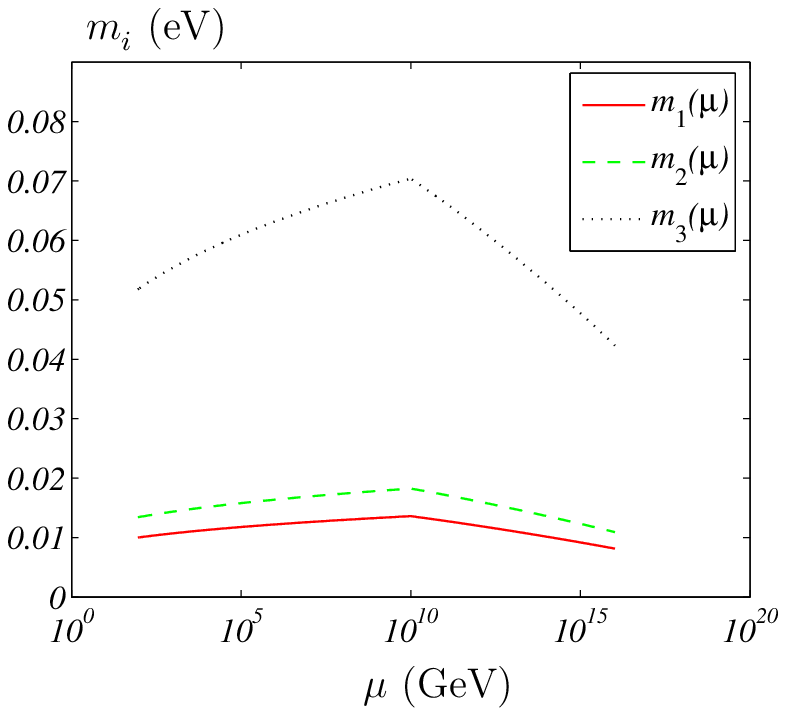, bbllx=1cm, bblly=6.0cm, bburx=11cm, bbury=16.0cm,%
width=11cm, height=11cm, angle=0, clip=0}\vspace{-2cm}\caption{The
running behaviors of light neutrino masses. Here we choose
$m^{}_{1}(M^{}_{Z})=0.01~{\rm eV}$ and the normal mass hierarchy
$m^{}_{1} < m^{}_{2} < m^{}_{3}$. We take the same value of
$\lambda^{}_H(M^{}_Z)$ as that in Fig. 2 in our calculations.}
\end{figure}


\begin{thebibliography}{99}
\bibitem{exp1} Super-Kamiokande Collaboration,
Y. Fukuda {\it et al.}, Phys. Rev. Lett. {\bf 81}, 1562 (1998);
Phys. Rev. Lett. {\bf 86}, 5656 (2001).

\bibitem{exp2} SNO Collaboration, Q.R. Ahmad {\it et al.},
Phys. Rev. Lett. {\bf 87}, 071301 (2001); Phys. Rev. Lett. {\bf
89}, 011302 (2002).

\bibitem{exp3} KamLAND Collaboration, K. Eguchi {\it et al.},
Phys. Rev. Lett. {\bf 90}, 021802 (2003).

\bibitem{exp4} CHOOZ Collaboration, M. Apollonio {\it et al.},
Phys. Lett. B {\bf 420}, 397 (1998); Palo Verde Collaboration, F.
Boehm {\it et al.}, Phys. Rev. Lett. {\bf 84}, 3764 (2000).

\bibitem{exp5} K2K Collaboration, M.H. Ahn {\it et al.},
Phys. Rev. Lett. {\bf 90}, 041801 (2003).

\bibitem{SS} P. Minkowski, Phys. Lett. B {\bf 67}, 421 (1977);
T. Yanagida, in {\it Proceedings of the Workshop on Unified Theory
and the Baryon Number of the Universe}, edited by O. Sawada and A.
Sugamoto (KEK, Tsukuba, 1979), p. 95; M. Gell-Mann, P. Ramond and
R. Slansky, in {\it Supergravity}, edited by F. van Nieuwenhuizen
and D. Freedman (North Holland, Amsterdam, 1979), p. 315; S.L.
Glashow, in {\it Quarks and Leptons}, edited by M.
L$\rm\acute{e}vy$ {\it et al.} (Plenum, New York, 1980), p. 707;
R.N. Mohapatra and G. Senjanovi\c{c}, Phys. Rev. Lett. {\bf 44},
912 (1980).

\bibitem{triplet}
G. Lazarides, Q. Shafi and C. Wetterich, Nucl. Phys. B {\bf 181},
287 {1981}; G.B. Gelmini and M. Roncadelli, Phys. Lett. B {\bf
99}, 411 (1981); E. Ma and U. Sarkar, Phys. Rev. Lett. {\bf80},
5716 (1998); I. Dorsner and P.F. Perez, Nucl. Phys. B {\bf 723},
53 (2005); T. Hambye, M. Raidal and A. Strumia, Phys. Lett. B {\bf
632}, 667 (2006); F.R. Joaquim and A. Rossi, Phys. Rev. Lett. {\bf
97}, 181801 (2006).

\bibitem{first}
A. Strumia and F. Vissani, {\it Neutrino masses and mixings
and...}, hep-ph/0606054.

\bibitem{GZZ}
N. Haba, N. Okamura and M. Sugiura, Prog. Theor. Phys. {\bf 103},
367 (2000); N. Haba, Y. Matsui, N. Okamura and M. Sugiura, Eur.
Phys. J. C {\bf 10}, 677 (1999); T. Fukuyama and N. Okada, JHEP
{\bf 0211}, 011 (2002); P.H. Gu, H. Zhang and S. Zhou, Phys. Rev.
D {\bf 74}, 076002 (2006); E.Kh. Akhmedov and M. Frigerio, JHEP
{\bf 01}, 043 (2007).


\bibitem{PDG}
Particle Data Group, W.M. Yao {\it et al}., J. Phys. G {\bf 33}, 1
(2006).

\bibitem{running}
P.H. Chankowski and Z. Pluciennik, Phys. Lett. B {\bf 316}, 312
(1993); K.S. Babu, C.N. Leung and J. Pantaleone, Phys. Lett. B
{\bf 319}, 191 (1993); S. Antusch, M. Drees, J. Kersten, M.
Lindner and M. Ratz, Phys. Lett. B {\bf 519}, 238 (2001); Phys.
Lett. B {\bf 525}, 130 (2002); S. Antusch, J. Kersten, M. Lindner,
M. Ratz and M.A. Schmidt, JHEP {\bf 0503}, 024 (2005); J.W. Mei,
Phys. Rev. D {\bf 71}, 073012 (2005); S. Luo, J.W. Mei and Z.Z.
Xing, Phys. Rev. D {\bf 72}, 053014 (2005); Z.Z. Xing, Phys. Lett.
B {\bf 633}, 550 (2006); Z.Z. Xing and H. Zhang, hep-ph/0601106.

\bibitem{lagranglian}
J.F. Gunion, R. Vega and J. Wudka, Phys. Rev. D {\bf 42}, 1673
(1990).

\bibitem{C}
R. Godbole, B. Mukhopadhyaya and M. Nowakowski, Phys. Lett. B {\bf
352}, 388 (1995).

\bibitem{integrating}
S. Weinberg, Phys. Lett. B {\bf 91}, 51 (1980); K.G. Wilson and
J.G. Kogut, Phys. Rep. {\bf 12}, 75 (1974); M. Bilenky and A.
Santamaria, Nucl. Phys. B {\bf 420}, 47 (1994).

\bibitem{RGE}
T.P. Cheng, E. Eichten and L.F. Li, Phys. Rev. D {\bf 9}, 2259
(1974); M. Machacek and M. Vaughn, Nucl. Phys. B {\bf 236}, 221
(1984).

\bibitem{para}
Z.Z. Xing, Int. J. Mod. Phys. A {\bf 19}, 1 (2004).

\bibitem{fix}
J.A. Casas, J.R. Espinosa, A. Ibarra and I. Navarro, Nucl. Phys. B
{\bf 573}, 652 (2000); S. Antusch, J. Kersten, M. Lindner and M.
Ratz, Nucl. Phys. B {\bf 674}, 401 (2003); S. Luo and Z.Z. Xing,
Phys. Lett. B {\bf 637}, 279 (2006).



\end{thebibliography}
\end{document}